\documentclass[aps,pre,twocolumn,showpacs,preprintnumbers,amsmath,amssymb,10pt,a4paper]{revtex4}
\makeatletter
\let\@ORGREVTEXendnotemark\@endnotemark
\let\@ORGREVTEX@makefnmark@cite\@makefnmark@cite
\def\@endnotemark{\bgroup\@fileswfalse\@ORGREVTEXendnotemark\egroup}
\def\@makefnmark@cite{\bgroup\@fileswfalse\@ORGREVTEX@makefnmark@cite\egroup}
\makeatother
\usepackage{geometry}
\usepackage{graphicx}
\usepackage{dcolumn}
\usepackage{bm}
\usepackage{subfigure}
\usepackage{amsmath}
\usepackage{amsfonts}
\usepackage{multirow}

\newcommand{\be}{\begin{equation}}
\newcommand{\ee}{\end{equation}}
\newcommand{\bd}{\begin{displaymath}}
\newcommand{\ed}{\end{displaymath}}
\newcommand{\BE}{\begin{eqnarray}}
\newcommand{\EE}{\end{eqnarray}}

\newcommand{\bmu}{\bar{\mu}}
\newcommand{\bx}{\bar{x}}

\newcommand{\bs}{\bm{s}}
\newcommand{\tx}{\widetilde{x}}

\newcommand{\txi}{\widetilde{\xi}}

\newcommand{\la}{\langle}
\newcommand{\ra}{\rangle}

\newcommand{\pd}[2]{\frac{\partial #1}{\partial #2}}

\begin{document}

\title{Learning to play public good games}

\author{Alex J. Bladon}
\email{alex.bladon@postgrad.manchester.ac.uk}
\author{Tobias Galla}
\email{tobias.galla@manchester.ac.uk}

\affiliation{Theoretical Physics, School of Physics and Astronomy, University 
of Manchester, Manchester M13 9PL, United Kingdom}

\date{\today}

\begin{abstract}
We extend recent analyses of stochastic effects in game dynamical learning to cases of multi-player games, and to games defined on networked structures. By means of an expansion in the noise strength we consider the weak-noise limit, and present an analytical computation of spectral properties of fluctuations in multi-player public good games. This extends existing work on two-player games. In particular we show that coherent cycles may emerge driven by noise in the adaptation dynamics. These phenomena are not too dissimilar from cyclic strategy switching observed in experiments of behavioural game theory.
\end{abstract}

\maketitle


\section{Introduction}
\label{sec:intro}

The theory of evolutionary dynamics is a cornerstone of modern biology. It has helped to describe a variety of systems where distinct elements compete to reproduce, for example in zoology and population dynamics~\cite{MaynardSmith}, but also in the dynamics of cancer growth and the different possible progressions of HIV~\cite{Nowakbook}, or in the evolution of language~\cite{Nowak2001}. Since the work of Maynard Smith~\cite{MaynardSmith} evolutionary ideas have also been applied in the context of game theory, and have augmented the more traditional approach to strategic decision making, based on equilibrium concepts~\cite{Neumann1944, Nash1950, Nash1951}. There is now an established field referred to as `evolutionary game theory' (see e.g. \cite{Gintis2000,Hofbauer1998}). This discipline is concerned with the study of populations of players, who interact in a game and who each carry a strategy and resulting fitness as a result of their success or otherwise in the game. Successful individuals then reproduce faster than those who are less successful in playing the game. Strategies are passed on from parent to offspring, with or without mutation, and the strategic content of the population of individuals hence evolves in time. Traditionally such processes have been modelled by means of replicator equations (or similar dynamical systems). These descriptions are typically deterministic in nature, and systematically neglect stochastic effects. More recently, the role of stochasticity and intrinsic noise has been considered in more detail, and the development of a mathematical theory with which to systematize these effects is very much work in progress. Phenomena brought about purely by stochasticity, and hence not captured by deterministic approaches, include for example fixation and the loss of biodiversity~\cite{Antal2006, Altrock2009}, drift reversal~\cite{Traulsen2004, Claussen2007}, or coherent stochastic oscillations~\cite{Bladon2010, Mobilia2010, Reichenbach2006}.

With an increasing number of experiments involving real human players being performed in laboratories of behavioural economics, ideas from evolutionary game theory have also been applied to describe the adaptation processes of game learning, see for example \cite{Traulsen2010, Hauert2002, Sato2002}. In this context, the strategic choices themselves are `genetic' elements. A `reproduction event' corresponds to an instance in which a human decides to switch from one strategy to another. These processes are frequently described mathematically by models drawn from evolutionary game theory. The advantage of using evolutionary game theory rests in the fact that there is a vast body of theoretical work available,  ready to be tied to experiments of behavioural game theory.  This approach does have its limitations though. Models of evolutionary game theory describe {\em populations} of agents, subject to selection pressure and generally involving birth and death processes. For obvious reasons no such birth-death dynamics takes place in a behavioural economist's laboratory, instead a number of fixed individuals interacts repeatedly in such experiments, without specific individuals being removed or replaced, born or killed.

It may therefore seem more appropriate to model the psychological decision making processes directly on the level of the interacting individual, rather than on a population level. Examples of such models can be  found in~\cite{Sato2002, Sato2002b,Camerer2003, Ho2007, Macy2002,Young2004,Fudenbergbook}. As one main ingredient of these psychologically motivated models players keep so-called `attractions' or `propensities' for each of the possible actions. These are then converted into a mixed strategy profile, and specific moves (pure actions) are played with the corresponding frequencies. Payoffs are received, depending on one's own move as well as on those made by the relevant opponents. In response to the outcome, players then update their propensities, increasing those of strategies that would have been successful against the observed moves, and reducing those of less successful actions. The process then iterates.

On the mathematical level, models of such processes define stochastic dynamical systems in discrete time. These models are not too dissimilar from learning algorithms studied in machine learning, including for example dynamical processes such as fictitious play and derivatives~\cite{Brown1951, Fudenberg1993,Fudenbergbook}. Stochasticity in learning can here have profound effects, depending on the details of the learning model adaptation can be seen to converge to Nash equilibria, or the learning process may fail to reach a stationary point due to persistent fluctuations and noise. 

Existing work on the analytical characterisation of noisy trajectories in game learning \cite{Galla2009,Galla2011} has revealed close similarities with stochastic effects in evolutionary processes in finite populations, but differences have been found as well \cite{Gomez2011}. Analytical studies have mostly been limited to the case of two-player learning i.e., games in which two individuals interact repeatedly and adapt to each other's moves. The correlation properties of fluctuations about limiting deterministic trajectories have been computed based on an expansion in the inverse noise strength, in good agreement with numerical simulations. The purpose of the present paper is to extend these studies and to address multi-player learning and learning of agents arranged on a (fixed) network. Stochastic effects in network learning have been considered before~\cite{Gomez2011}, but no systematic analytical characterisation has been attempted. As one result of our analysis we will demonstrate that amplified stochastic oscillations, commonly found in models of evolutionary game theory, also appear in multi-player and network learning.  We also analyse how the networks structure and parameters of the earning rule affect the outcome of adaptation.

 The remainder of the paper is organised as follows: in Sec. \ref{sec:gamelearning} we outline a simple model of reinforcement learning and define the public good game we will be studying subsequently. Sec. \ref{sec:deterministic} contains a brief analysis of a deterministic limiting case of the learning process. We then classify the oscillations associated with stochastic multi-player learning in Sec. \ref{sec:stoc}. We analyse the resulting power spectra and discuss their implications for game learning in Sec. \ref{sec:results}. Network games are then considered  in Sec. \ref{sec:networks}. Sec. \ref{sec:conclusion} contains concluding remarks and an outlook on possible future lines of research.
\section{Model and definitions}
\label{sec:gamelearning}
In this section we define the public goods game and the learning algorithm which we will analyze in the subsequent sections.
\subsection{Public good game}

In a typical public good game each player decides whether or not to contribute an amount $c$ to a shared `pot', for example a contribution to a common effort. The amount in the pot is then multiplied by a factor $r>1$ and re-distributed amongst all players involved in the game, no matter whether they contributed or not. If a player does not contribute they still receive a share of the pot, however the fewer people contribute the smaller each individual's share becomes. Thus the public goods game can be thought of as a multi-player social dilemma, similar to the celebrated two-person prisoner's dilemma. There is a temptation to defect (i.e., not to contribute), but the reward for a group consisting only of defectors is less than that for a group of contributors. In the public goods game outlined by \cite{Hauert2002} this basic setup is extended to allow players not to participate in the game at all. Such players, called `loners', instead receive a small, but guaranteed payout of $\sigma c$, where $\sigma$ is some non-zero constant, chose such that $\sigma c$ is less than $(r-1)c$. This produces a cyclic relationship between the three actions: if everyone is contributing the best thing to do is defect; if everyone is defecting, playing `loner' is the best option and if everyone is a loner then the greatest payoff comes from contributing.
 
The net payoff for each strategy (defined as the difference between the money received at the end and the money put in at the beginning) can be written as

\BE\label{eq:payoff}
\pi_C &=& -c + rc\frac{n_C+1}{1+n_C+n_D}, \\ \nonumber
\pi_D &=& rc\frac{n_C}{1+n_C+n_D}, \\ \nonumber
\pi_L &=& \sigma c,
\EE

where $n_C$ and $n_D$ are numbers of contributors and defectors in the \em rest \rm of the group interacting in a particular instance of the game. In other words, the above payoff $\pi_C$ is that of a co-operator playing with a group of $n_C$ {\em other} co-operators and $n_D$ defectors, and similar for $\pi_D$. At variance with some definitions of public good games, we here assume that the above payoff relations also hold if only one single player decides to participate in the game, and if all other players decide to abstain. This choice is mainly made for analytical convenience. Unless otherwise specified we will use $r=1.8$, $c=1$ and $\sigma = 0.5$ (as in \cite{Hauert2002}). We will abbreviate the three pure strategies as $C$, $D$ and $L$ in the following.

\subsection{Reinforcement learning}\label{sec:reinf}
We consider the game learning model proposed by \cite{Sato2002}. In such models, $N$ players play a game attempting to maximise their payoffs. In each round of the game each player uses one of the $S$ available pure strategies, each with a probability defined by the strategy's previous performance as perceived by the player. This will be specified further below. For the case of the public goods game we have $S=3$ pure strategies, contribute (C), defect (D) and `loner' (L), but our theory is applicable for general $S$. 
\subsubsection{Update of propensities}
The probabilities with which a given player chooses to play the three actions form a mixed strategy profile in the language of game theory. In our learning model the strategy profile will be based on so-called `scores' (or propensities) that each player assigns to the pure strategies available to him/her. These scores measure the performance of the pure actions against the observed actions of opponents.  The score given to strategy $s\in\{1,\dots,S\}$ by player $i$ is assumed to be updated once every $\Omega$ rounds of the game, and kept constant inbetween. Real-world play corresponds to $\Omega=1$, i.e. updating after each round. For our purposes it is however useful to introduce the more general dynamics as (see also \cite{Galla2009,Galla2011,Sato2005})
 \be
q_{i,s}(t+\Omega) = (1 - \lambda)q_{i,s}(t) + \frac{1}{\Omega}\sum_{t^\prime = t}^{t^\prime = t + \Omega - 1}u(s, \bm{s}_{-i}(t^\prime)),
\label{eqn:score}
\ee
where $q_{i,s}$ is assumed to remain unchanged between time $t$ and time $t+\Omega-1$. Here $u(s,\bm{s}_{-i}(t))$ is the payoff for player $i$ when they play strategy $s$ at time $t$ given that the other players play the pure strategies denoted by $\bs_{-i}\in\{1,\dots,S\}^{N-1}$. In our case $s$ can take the values $s\in\{\mbox{C,D,L}\}$, and similarly each entry in $\bs_{-i}$ takes values in $\{\mbox{C,D,L}\}$. The payoff $u(s,\bm{s}_{-i}(t))$ is given by expressions (\ref{eq:payoff}), where $n_C$ and $n_D$ are the number of entries $C$ and $D$ respectively in the vector $\bm{s}_{-i}$.  Some more explanation of the update rule of Eq. (\ref{eqn:score}) is here appropriate. The above learning rule assumes that adaptation occurs only once every $\Omega$ iterations of the game. This corresponds to what is known as batch learning \cite{Saadbook}. The last term on the RHS of Eq. (\ref{eqn:score}) is indeed the {\em average} payoff per round obtained by player $i$ in $\Omega$ iterations of the game, and given his or her opponents' actions $\bs_{-i}(t')$ during those $\Omega$ rounds ($t'=t,t+1,\dots,t+\Omega-1$). The philosophy behind introducing batch-learning dynamics is discussed further below in the context of the deterministic limit of the stochastic dynamics (Sec. \ref{sec:det}). The pre-factor $1-\lambda$ in the first term on the RHS of the update rule describes memory loss. For $\lambda=0$ the score $q_{i,s}(t)$ is proportional to the payoff player $i$ would have received up to time $t$ had he or she played action $i$ at all times, and given his or her opponents' actions. All past play carries equal weight and is accumulated. The parameter $\lambda\in[0,1]$ is a memory-loss parameter. For $\lambda>0$ past play is exponentially discounted, and observations in the distant past carry a lesser weight than more recent iterations of the game. Similar mechanisms are present in learning models of behavioural game theory, see for example~\cite{Camerer2003}, they have also been used in~\cite{Sato2005}. Occasionally a pre-factor $\lambda$ is introduced in the second term on the RHS of Eq. (\ref{eqn:score}). This is mainly a matter of notation, and amounts to a re-scaling of payoffs. We here use the notation of \cite{Sato2002,Sato2005}.

\subsubsection{Fully connected and networked setups}
In writing down Eq. (\ref{eqn:score}) we have implicitly assumed that the payoff of any given player $i$ may at least potentially depend on the actions $\bs_{-i}$ of {\em all} other players. This corresponds to a fully connected (or well-mixed) population, in which each player interacts with any other player. Specifically, a well mixed model describes a group of $N$ players, who {\em all} engage in the same $N$-player public good game by each choosing from the three actions $\{C,D,L\}$. Different setups have been considered for example in \cite{Pacheco2009} where a group of $N$ players is selected randomly from a larger populations of $Z$ individuals.

We also consider the case of a networked arrangement of players. Here agents are placed on the nodes of a graph, and each player is then assumed to repeatedly play iterations of the public goods game with their neighbours on the graph. Evolutionary descriptions of such network public goods games have been considered for example in~\cite{Santos2008}. The analog of the adaptation rule of Eq. (\ref{eqn:score}) is then given by
 \be
q_{i,s}(t+\Omega) = (1 - \lambda)q_{i,s}(t) + \frac{1}{\Omega}\sum_{t^\prime = t}^{t^\prime = t + \Omega - 1}u(s, \bm{s}_{\partial i}(t^\prime))
\label{eqn:score2},
\ee
where $\partial i$ denotes the set of neighbours of $i$ in the network, and where $\bs_{\partial i}$ accordingly is the vector of pure actions taken by those neighbours. The learning dynamics on networks will be discussed in more detail below (see Sec. \ref{sec:networks}).
\subsubsection{Conversion into mixed strategies}
In our model the scores (or propensities) $\{q_{i,s}\}$ determine the mixed strategy profiles of players using the so-called logit (or softmax) rule
\be
x_{i,s}(t) = \frac{e^{\beta q_{i,s}(t)}}{\sum_{s^\prime} e^{\beta q_{i,s^\prime}(t)}},
\label{eqn:logit}
\ee
where $x_{i,s}(t)$ is the probability of player $i$ using strategy $s$ in round $t$.  This scheme has similar features to the Fermi update rule commonly used in evolutionary dynamics \cite{Traulsen2007, Bladon2010, Pacheco2006}. Rules of this type have also been used to fit data from experiments with real players, see e.g.  \cite{Traulsen2010, Ho2007, Camererbook}.

The parameter $\beta$ in Eq. (\ref{eqn:logit}) is referred to as the intensity of choice (or learning rate) and determines how much importance is given to a difference in payoffs when calculating the mixed strategy profile. If $\beta=0$ for example the players choose their actions completely at random and disregard their propensities entirely. If $\beta=\infty$ then they strictly play the pure action with the highest score. 

Following~\cite{Sato2005} it is possible to combine Eq.~(\ref{eqn:logit}) and Eq.~(\ref{eqn:score}) into map 
\BE\label{eqn:stocmap}
&&x_{i,s}(t+\Omega) = \\ &&\frac{{x_{i,s}(t)^{1-\lambda}\exp (\frac{\beta}{\Omega}\sum_{t^\prime = t}^{t^\prime = t + \Omega - 1} u(s,\bm{s}_{-i}(t^\prime))}
)}{\sum_{s^\prime} x_{i,s^\prime}(t)^{1-\lambda}\exp (\frac{\beta}{\Omega}\sum_{t^\prime = t}^{t^\prime = t + \Omega - 1} u(s^\prime,\bm{s}_{-i}(t^\prime)))}. \nonumber
\EE
This equation defines the discrete-time evolution of the mixed strategy vectors of players, no reference to the propensities is required. Eq. (\ref{eqn:stocmap}) here applies to the well-mixed case, but it is straightforward to formulate the corresponding map for a networked system. 
\subsubsection{Deterministic limit}\label{sec:det}
Models in behavioural game theory \cite{Camerer2003, Ho2007} are typically based on frequent update of propensities and mixed strategy profiles. Adaptation is assumed to occur after each iteration of the game, corresponding to $\Omega=1$ in our model. Such dynamics are intrinsically stochastic, as moves of all players are drawn at random from the underlying mixed strategy profiles. The main purpose of the present work is to investigate the effects this stochasticity has on the outcome of learning. It is however generally very difficult to obtain analytical results for stochastic dynamical systems such as the one corresponding to the case of $\Omega=1$. We therefore follow an approach similar to that used in evolutionary dynamics: (i) we first derive the {\em deterministic} limit dynamics. This corresponds to taking the limit $\Omega\to\infty$, and is akin to considering the limit of {\em infinite} populations in evolutionary game theory (resulting in deterministic differential equations of the replicator or replicator-mutator type), (ii) we then perform a systematic perturbative expansion in $\Omega^{-1/2}$ in order to capture stochastic effects to lowest non-trivial order. This is again similar to approaches taken in the context of evolutionary games, where expansions in the inverse population-size are often carried out, see for example \cite{Traulsen2004,Claussen2007,Bladon2010, Mobilia2010, Reichenbach2006}. While such expansions in the inverse batch size are technically only valid for large, but finite $\Omega$ we will show below that they can capture the on-line dynamics, $\Omega=1$ in good approximation as well.

We will here briefly derive the deterministic limit of learning, and analyze the outcome in the next section. The expansion in the noise strength is then carried out in Sec. \ref{sec:stoc}.

In the limit $\Omega \rightarrow \infty$, Eq.~(\ref{eqn:stocmap}) the mean of the payoffs from the last $\Omega$ rounds approaches the expected payoff for player $i$ when using a given pure strategy, where the expectation is to be taken with respect to the mixed strategies of $i$'s opponents. Specifically we have
\BE
&&\lim_{\Omega\to\infty}  \frac{1}{\Omega}\sum_{t^\prime = t}^{t^\prime = t + \Omega - 1}u(s, \bm{s}_{-i}(t^\prime))\nonumber \\
&=&\mu_{i,s} \equiv \sum_{\bm{s}_{-i}}u(s,\bm{s}_{-i})p_{-i,\bm{s}_{-i}},
\EE
where $p_{-i,\bs_{-i}}=\prod_{j\neq i} x_{j,s_j}$ is the probability of player $i$ facing the pure strategies $\bs_{-i}\in\{1,\dots,S\}^{N-1}$ being played by his or her opponents. The sum over $\bm{s}_{-i}$ in the above expression extends over all $S^{N-1}$ possible realizations of the vector $\bs_{-i}$.
Rescaling time by a suitable factor the update rule of Eq. (\ref{eqn:stocmap}) then becomes
\be
x_{i,s}(\tau+1) = \frac{x_{i,s}(\tau)^{1-\lambda}\exp (\beta\mu_{i,s}(\tau))
)}{\sum_{s^\prime} x_{i,s^\prime}(\tau)^{1-\lambda}\exp (\beta\mu_{i,s^\prime}(\tau))},
\label{eqn:detmap}
\ee
Again, this deterministic map was formulated for the well-mixed case, generalization to network learning is straightforward.

It is also possible to write Eq.~(\ref{eqn:logit}) and Eq.~(\ref{eqn:score}) together in the form of an approximate continuous time dynamic if $\beta \ll 1$. This is known as the Sato-Crutchfield dynamics \cite{Sato2005}. These dynamics are similar to the replicator-mutator equations where the learning rate is analogous to selection strength and the memory loss is analagous to the mutation rate \cite{Nowak2001a,Bladon2010}. We will restrict our analysis to the map as we feel it better represents the discrete rounds used in experiments.

\section{The fully connected case: Results of the deterministic analysis}
\label{sec:deterministic}
In this section we will first focus on the fully connected (or well-mixed) setup, networked systems are addressed below. Before discussing the outcome of deterministic learning is it useful to briefly comment on the role of initial conditions. If the mixed strategy profiles for the different players are initiated from the same point in the space of all mixed strategies i.e., $x_{i,s}(\tau=0)=x_{j,s}(\tau=0)$ for all players $i,j$ and all pure actions $s$, then the deterministic map, Eq. (\ref{eqn:detmap}), will preserve this symmetry, i.e., the mixed strategy profiles will remain identical across players, $x_{i,s}(\tau)=x_{j,s}(\tau)$ for later times $\tau$ as well. We will refer to such cases as a start from homogeneous inital conditions. Each player will then receive the same average payoffs in each round, and the strategy vectors of all players will evolve identically. Thus each player is effectively playing with $N-1$ copies of themselves. For heterogeneous initial conditions, or noisy dynamics, this is not necessarily the case.

\begin{figure}
\vspace{2.5em}
\centering
\includegraphics[scale=0.4]{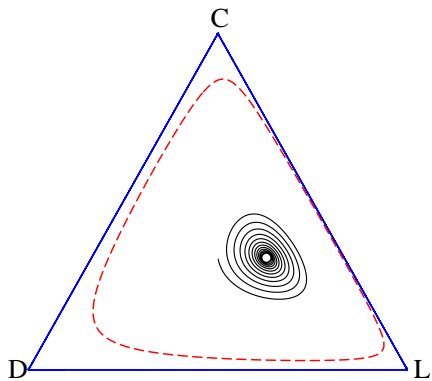}
\caption{\label{fig:dynamics} (Color online) Dynamics of the three-player public good game with homogeneous initial conditions (intensity of choice is $\beta = 0.1$). The black line indicates the trajectory to the fixed point at $\lambda = 10^{-3}$, the dashed red line is the limit cycle at $\lambda=10^{-4}$ and the solid blue line is the limit cycle around the edge of the simplex at $\lambda = 10^{-7}$.}
\end{figure}

We will begin by examining the behaviour of the deterministic system, Eq.~(\ref{eqn:detmap}), for homogeneous initial conditions. The deterministic dynamics is found to exhibit three types of behaviour: (i) for sufficiently large values of the memory-loss parameter $\lambda$ trajectories approache a stable fixed point, (ii) for intermediate values of $\lambda$ one finds convergence towards limit cycles in the interior of strategy space, and  (iii) at small memory-loss asymptotic trajectories along the edges of the strategy simplex are seen. Examples are shown in Fig.~\ref{fig:dynamics}.

\begin{figure}
\vspace{2.5em}
\centering
\includegraphics[scale=0.5]{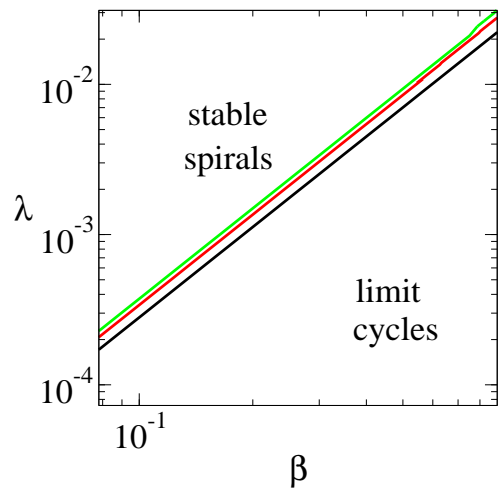}
\caption{\label{fig:phase} (Color online) Phase diagram characterizing the outcome of deterministic learning in the $N$-player public goods game. Lines show the location of the Hopf bifurcation in parameter space, separating stable from unstable fixed points. From bottom to top, lines corresponds to $N=3$ (black), $N=4$ (red) and $N=5$ players (green).}
\end{figure}

\begin{figure}
\vspace{2.5em}
\centering
\includegraphics[scale=0.5]{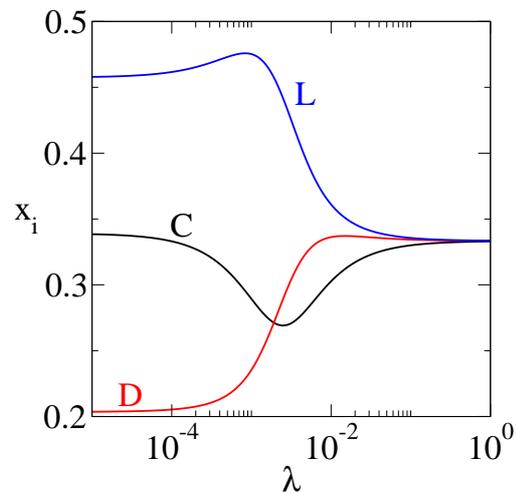}
\caption{\label{fig:fixedpoints} (Color online) This figure shows the components of the fixed point as a function of the memory loss parameter $\lambda$. Here, $N=3$ and $\beta = 0.01$. The black line shows the concentration of contribution in the mixed strategy profile, red the concentration of defection and green the concentration of the loner strategy. }
\end{figure}

A more systematic analysis can be found in Fig.~\ref{fig:phase} where we show phase diagrams for games of $N=3$, $4$ and $5$ players. The location and existence of the fixed point is determined numerically and, as such, the boundary between stable spirals and limit cycles reported in the figure are approximate. Performing a linear stability analysis of the fixed points reveals that stable fixed points are always found to be spirals (the eigenvalues of the Jacobian form complex conjugate pairs with negative real parts). The transition to the limit cycle regime occurs through a Hopf bifurcation. The size and degree of stability of the limit cycles is determined by the distance to the Hopf bifurcation in the parameter space. As this distance increases the limit cycles move closer to the border of the simplex, eventually becoming restricted to the border itself (as seen in Fig.~\ref{fig:dynamics}).
The dynamics of the limit cycles follow a periodic pattern, with contributing, defecting and loner each becoming the most prominent strategy in turn. The effect of the memory loss parameter on the fixed point of the players is shown in Fig.~\ref{fig:fixedpoints}, at low memory loss the players tend to  mostly abstain (L), at large memory loss play occurs essentially at random, with all three actions being used with nearly the same frequencies.

We will mostly ignore the case of heterogeneous initial conditions in the following, in all tested cases we find numerically that the system approaches the same fixed point as for homogeneous initial conditions. Since we will be interested in features where the deterministic system is at a fixed point, restricting our investigations to the simpler case of homogeneous initial conditions will not affect our findings.

\section{Batch-size expansion for the well-mixed case}
\label{sec:stoc}

\begin{figure}
\vspace{2.5em}
\centering
\includegraphics[scale=0.5]{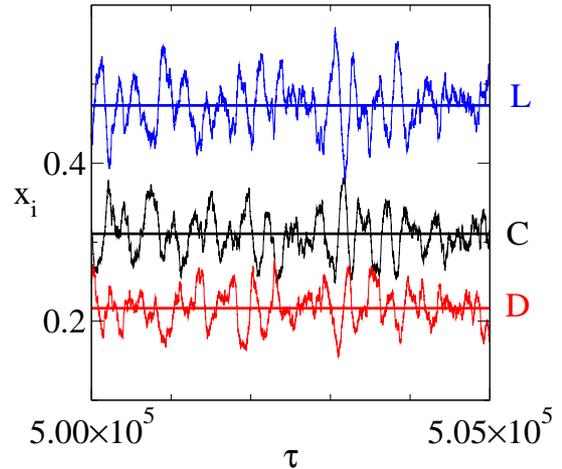}
\caption{An example of the stochastic oscillations in the mixed strategy profiles of a fixed player for $\beta = 0.1$ and $\lambda = 0.005$. The three oscillating curves show the mixed strategy obtained from one simulation run of the stochastic learning dyamics (L, C and D from top to bottom). Solid lines indicate the deterministic fixed point. }
\label{fig:stoc}
\end{figure}

Similar to what is seen in population-based models of evolutionary game theory  \cite{Bladon2010, Traulsen2008, Traulsen2009,Mobilia2010, Reichenbach2006}, models of epidemics \cite{Black2009, Alonso2007, Coulson2004, Simoes2007} and in other biological systems~\cite{McKane2005, Scott2006, Gonze2002} the behaviour of the stochastic dynamics of learning can be quite different from that predicted by the deterministic description. In particular a mechanism of coherent amplification can lead to sustained stochastic cycles in the noisy system for values of the model parameters in which deterministic learning converges.  Fig.~\ref{fig:stoc} shows an example of this behaviour. In population-based models these oscillations are known as `quasi-cycles', they are an effect seen in finite populations, caused by so-called `demographic noise'~\cite{McKane2005}, noise induced by sampling interacting agents from a finite pool of individuals.  Deterministic descriptions of evolving systems are valid only in the limit of {\em infinite} populations.  In the context of two-player learning similar cycles have been studied in \cite{Galla2009, Galla2011}. While the mechanism of stochastic amplification is similar to that in population dynamics, the source of noise is different. As described in Sec.~\ref{sec:reinf} the players' adaptation is intrinsically stochastic if they base their strategy updates on {\em finite} sets of observations of their opponents' moves. In Eq.~(\ref{eqn:score}) for example we have assumed that a finite number $\Omega$ of observations is made between any two adaptation events. Learning becomes deterministic only in the limit $\Omega\to\infty$. In this sense the batch size $\Omega$ is similar to the population size in evolving systems, they both control the strength of noise in the resulting dynamics. This observation is one of the main reasons for introducing the learning algorithm at general batch size, tuning $\Omega$ allows us to interpolate between the realistic case of frequent updating ($\Omega=1$), and the deterministic limit, $\Omega\to\infty$.

We will now proceed to obtain an analytical characterization of the stochastic quasi-cycles shown in Fig. \ref{fig:stoc}. To this end we carry out a systematic expansion in the noise strength, more precisely in powers $\Omega^{-1/2}$ of the inverse batch size. This is conceptually similar to the van Kampen expansion~\cite{Kampenbook} in powers of the inverse system size of population-based models. We here discuss the key steps of the calculation for the general $N$-player game, choosing a suitably compact notation. In order to make the mathematical details more transparent we also detail the resulting expressions for the explicit case $N=3$ in the Appendix. The first step of the expansion is to rewrite the last term on the RHS of Eq.~(\ref{eqn:score}). This term represents the average payoff per round to player $i$ in $\Omega$ iterates of the game. Separating stochastic fluctuations from the expected {\em mean} payoff this term can be written as follows
\be
\frac{1}{\Omega}\displaystyle\sum_{t' = t}^{t'=t+\Omega-1} u(s,\bm{s}_{-i}(t^\prime)) = \mu_{i,s} + \widetilde{\xi}_{i,s}
\label{eqn:payexp}
\ee
where $\mu_{i,s}=\sum_{\bm{s}_{-i}}u(s,\bm{s}_{-i})p_{-i,\bm{s}_{-i}}$ is the expected payoff (per round) for player $i$ given his opponents' mixed strategy profiles, and provided player $i$ plays pure strategy $s$. The second contribution, $\widetilde{\xi}_{i,s}$, represents the fluctuations about the deterministic trajectory. The correlations between these noise terms are found by rearranging Eq.~(\ref{eqn:payexp}), isolating $\txi_{i,s}$ and averaging over all possible actions of all players. Correlations between noise variables associated with strategies belonging to the same player are given by
\BE\label{eqn:samecorr}
\la\txi_{i,s}(\tau)\txi_{i,s^\prime}(\tau')\ra &=& \frac{\delta_{\tau \tau^\prime}}{\Omega}\sum_{\bm{s}_{-i}}(u(s,\bm{s}_{-i}) - \mu_{i,s})\\&&\times(u(s^\prime,\bm{s}_{-i}) - \mu_{i,s^\prime})p_{-i,\bs_{-i}}. \nonumber
\EE
Unlike in the two-player case \cite{Galla2009, Galla2011}, there are now correlations also between the noise variables associated with different players. This is because the payoffs of any two players both depend on the actions of the remaining players. These correlations take the form
\BE\label{eqn:diffcorr}
\la\txi_{i,s}(\tau)\txi_{j,s^\prime}(\tau')\ra &=& \\  &&\frac{\delta_{\tau \tau^\prime}}{\Omega}\sum_{\bm{s}}(u(s,\bm{s}_{-i}) - \mu_{i,s})\nonumber\\&\times&(u(s^\prime,\bm{s}_{-j}) - \mu_{j,s^\prime})p_{\bs}, \nonumber
\EE
where $\bs=(s_1,\dots,s_N)\in\{1,\dots,S\}^N$ describes the vector of actions taken by the whole group of players, and where $p_{\bs} = \prod_{k}x_{k,s_k}$. It is worth pointing out that Eq. (\ref{eqn:samecorr}) can be seen as a special case of Eq. (\ref{eqn:diffcorr}), if $i=j$ in the latter relation then the sum over actions of player $i$ can be factored out, and one obtains Eq. (\ref{eqn:samecorr}).

Given that the payoffs, Eq.~(\ref{eqn:payexp}), are now random variables, the same will be true for the components $x_{i,s}$ of the players' mixed strategy profiles. In order to separate fluctuations from the deterministic dynamics we use the ansatz 
\be
x_{i,s}(\tau) = \bx_{i,s}(\tau) + \tx_{i,s}(\tau),
\ee 
where the first term on the RHS is the deterministic trajectory, and where the second term is of order $\Omega^{-1/2}$ and describes fluctuations about the deterministic solution.
Using this ansatz, together with Eq.~(\ref{eqn:payexp}), and after the re-scaling of time discussed above, we can rewrite Eq.~(\ref{eqn:stocmap}) as
\BE\label{eqn:start}
&&x_{i,s}(\tau+1) = \nonumber \\
&&\frac{(\bar{x}_{i,s} + \widetilde{x}_{i,s})^{1-\lambda}e^{\beta(\bar{\mu}_{i,s}+ \widetilde{\mu}_{i,s} + \widetilde{\xi}_{i,s})}}{\sum_{s^\prime}(\bar{x}_{i,s^\prime} + \widetilde{x}_{i,s^\prime})^{1-\lambda}e^{\beta(\bar{\mu}_{i,s^\prime} + \widetilde{\mu}_{i,s^\prime} + \widetilde{\xi}_{i,s^\prime})}},
\EE
where 
\be
\bar{\mu}_{i,s} = \sum_{\bm{s}_{-i}}u(s,\bm{s}_{-i})\bar{p}_{-i,\bm{s}_{-i}},
\ee
\be
\widetilde{\mu}_{i,s} =  \sum_{\bm{s}_{-i}}u(s,\bm{s}_{-i})\widetilde{p}_{-i,\bm{s}_{-i}},
\ee 
and where all quantities on the RHS of Eq. (\ref{eqn:start}) are evaluated at time $\tau$.
We have here introduced $\bar{p}_{-i,\bm{s}_{-i}} = \prod_{k\neq i}\bar{x}_{k,s_k}$ and 
\BE
\widetilde{p}_{-i,\bm{s}_{-i}} &=& \sum_{k\neq i} \frac{\partial \bar{p}_{-i,\bm{s}_{-i}}}{\partial \bar x_{k,s_k}}\widetilde{x}_{k,s_k}\nonumber \\
&=&\sum_{k\neq i}\left(\prod_{j\notin\{i,k\}} \bar{x}_{j,s_j}\right)\widetilde{x}_{k,s_k}.
\EE
We now expand Eq. (\ref{eqn:start}) in powers of $\Omega^{-1/2}$, taking into account that $\tx_{i,s}$ and $\txi_{i,s}$ scale as $\Omega^{-1/2}$ $\forall i,s$. Naturally, from the leading-order terms in this one expansion recovers the deterministic dynamics
\be
\bx_{i,s}(\tau+1)  = \frac{\bx_{i,s}^{1-\lambda}e^{\beta\bmu_{i,s}}}{\sum_{s^\prime}\bx_{i,s^\prime}^{1-\lambda}e^{\beta\bmu_{i,{s^\prime}}}}.
\ee
At next-leading order in the expansion one finds
\BE\label{eqn:order2}
\tx_{i,s}(\tau+1) &=& \sum_k\sum_{s^\prime}\left(\left.\pd{g_{i,s}}{\widetilde{x}_{k,s^\prime}}\right)\right|_{\widetilde {\bm x}=\widetilde{\bm \xi}=0}\widetilde{x}_{k,s^\prime} \nonumber \\
&&+ \sum_{s^\prime}\left(\left.\pd{g_{i,s}}{\widetilde{\xi}_{i,s^\prime}}\right)\right|_{\widetilde {\bm x}=\widetilde{\bm \xi}=0}\widetilde{\xi}_{i,s^\prime},
\EE
where $g_{i,s}$ is a short-hand for the RHS of Eq.~(\ref{eqn:start}). The notation $\widetilde {\bm x}=\widetilde{\bm \xi}=0$ indicates that all variables $\tx_{i,s}$ and $\txi_{i,s}$ ($i=1,\dots,N$, $s=1,\dots,S$) are to be taken to zero when evaluating the derivatives. The derivative of $g_{i,s}$ with respect to $\widetilde{\xi}_{k,s^\prime}$   vanishes for $k\neq i$. The resulting coefficients in Eq.~(\ref{eqn:order2}) multiplying the variables $\{\tx_{k,s'}\}$ are the entries of the Jacobian of Eq.~(\ref{eqn:start}), to be evaluated at $\widetilde x_{i,s}=\widetilde\xi_{i,s}=0$. We now further simplify these terms and find
\BE
&&\sum_{s^\prime}\left(\left.\pd{g_{i,s}}{\widetilde{\xi}_{i,s^\prime}}\right)\right|_{\widetilde {\bm x}
=\widetilde{\bm \xi}=0}\widetilde{\xi}_{i,s^\prime}\nonumber \\
&=& \beta\bx_{i,s}\left(\txi_{i,s} - \sum_{s^\prime}\bar{x}_{i,s^\prime}\txi_{i,s^\prime}\right),
\EE
similar to what has been reported in~\cite{Galla2011}. The resulting dynamics for the fluctuations $\widetilde x_{i,s}$ about the deterministic trajectory can be compactly written as a set of Langevin equations for the $N\times S$ variables, $\widetilde x_{i,s}$,
\be
\bm{\delta}(\tau+1) = \bm{\delta}(\tau) + \mathcal{J}\bm{\delta}(\tau) + \bm{\zeta}.
\label{eqn:langevin}
\ee
We have here introduced $\bm{\delta} = (\widetilde{\bm{x}}_1,\dots,\widetilde{\bm{x}}_N)$, where $\widetilde{\bm{x}}_i = (\widetilde{x}_{i,1},\dots,\widetilde{x}_{i,S})$. We have also written $\bm{\zeta} =(\widetilde{\bm{\gamma}_{1}},\dots,\widetilde{\bm{\gamma}_{N}})$ with $\widetilde{\bm{\gamma}}_i=(\gamma_{i,1},\dots,\gamma_{i,s})$, and where
\be
\gamma_{i,s} = \beta\bx_{i,s}\left(\txi_{i,s} - \sum_{s^\prime}\bx_{i,s^\prime}\txi_{i,s^\prime}\right).
\label{eqn:gamma}
\ee
The $(NS)\times(NS)$-matrix $\mathcal{J}$ is the Jacobian of Eq.~(\ref{eqn:start}), evaluated at the fixed point of the deterministic dynamics. The correlations between the components of ${\bf \gamma}$, can be expressed in terms of those of the $\{\widetilde\xi_{i,s}\}$, specifically we have
\BE
&&\langle\gamma_{i,s}\gamma_{j,s^\prime}\rangle \nonumber\\
&=& \beta^2\bx_{i,s}\bx_{j,s^\prime}\bigg( \langle\txi_{i,s}\txi_{j,s^\prime}\rangle - \bx_{i,s}\sum_{s''}\bx_{j,s''}\langle\txi_{i,s}\txi_{j,s''}\rangle \nonumber \\
&&- \bx_{j,s^\prime}\sum_{s''}\bx_{i,s''}\langle\txi_{i,s''}\txi_{j,s^\prime}\rangle \nonumber \\
&& + \sum_{s''s''''}\bx_{i,s''}\bx_{j,s'''}\langle\txi_{i,s''}\txi_{j,s'''}\rangle\bigg).
\EE
Fourier transforming Eq.~(\ref{eqn:langevin}) gives
\be
\left((e^{i\omega} - 1)\mathcal{I} - \mathcal{J}\right)\widehat{\bm{\delta}} = \widehat{\bm{\zeta}},
\ee
where $\widehat{\bm {\delta}}$ is the Fourier transform of ${\bm\delta}$ and $\widehat{\bm\zeta}$ that of $\bm{\zeta}$. Defining $\mathcal{M}(\omega) = \left((e^{i\omega} - 1)\mathcal{I} - \mathcal{J}\right)$, with $\mathcal{I}$ the identity matrix, and using the notation $\mathcal{B}$ for the matrix of correlations of $\bm \gamma$ we find
\be
\left<\widehat{\bm{\delta}}(\omega)\widehat{\bm{\delta}}(\omega')^T\right> = \mathcal{M}(\omega)^{-1}\mathcal{B}(\mathcal{M}^\dagger(\omega))^{-1}\delta(\omega+\omega').
\label{eqn:spectra}
\ee
We will refer to the matrix $P(\omega)=\mathcal{M}(\omega)^{-1}\mathcal{B}(\mathcal{M}^\dagger(\omega))^{-1}$ as the set of power spectra of fluctuations in the following. This is the final result of the batch-size expansion. Although formulated in Fourier space the expressions in Eq.~(\ref{eqn:spectra}) contain full information about the autocorrelation functions and crosscorrelation functions of the variables $\widetilde x_{i,s}$. They hence characterize the fluctuations about the deterministic fixed point (during the course of the calculation we have assumed that the deterministic trajectory approaches a stable fixed point, generalization to limit cycles is possible, but tedious, see \cite{Boland2008,Boland2009}).  It is straightforward to carry out the required matrix inversions, and to evaluate the RHS of Eq.~(\ref{eqn:spectra}) numerically. We will here mostly focus on quasi-cycles, it is therefore convenient to operate in Fourier space. In particular we will compare the power spectra of quasi-cycles, as obtained from Eq.~(\ref{eqn:spectra}) against numerical simulations in the next section. Given that we have carried out an expansion in powers of $\Omega^{-1/2}$ and that we have only retained the next-to-leading order we expect our results to be valid for large, but finite values of the batch size $\Omega$. 

\section{Characterization of quasi-cycles in the fully connected case}
\label{sec:results}

\begin{figure}
\vspace{3em}
\centering
\includegraphics[scale=0.5]{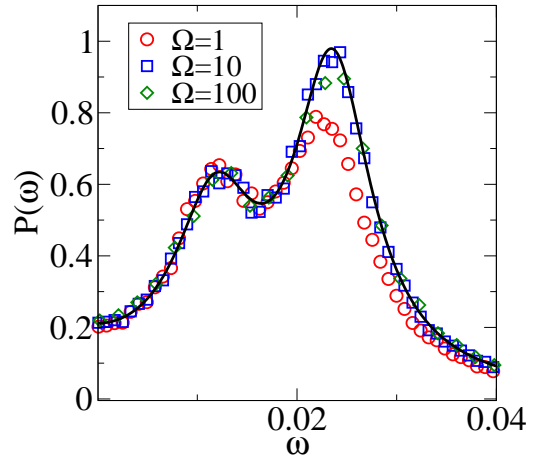}
\caption{\label{fig:ps_ocomp} This figure shows the power spectrum of the contributory strategy in a three player game at $\lambda = 0.005$, $\beta = 0.1$. The different symbols shows simulations at $\Omega = 1, 10, 100$, averaged over in excess of $100$ independent runs. The solid line is the theoretical spectrum from Eq.~(\ref{eqn:spectra}).}
\end{figure}

We now proceed to test the theoretical results obtained in the previous section against numerical simulations of the multi-player learning process. Taking a three-player game as an example, we find good agreement between simulation and theory, see Fig.~\ref{fig:ps_ocomp}. This figure also demonstrates the finite-size effects at different batch sizes. The case $\Omega = 1$ is the so-called `on-line' learning limit where players update their scores after every round~\cite{Saadbook}. Here the agreement with the theory is only approximate, which is no surprise given that higher-order terms in the expansion have been discarded. Nevertheless the anaytical theory is able to predict the dominating frequency of the quasi-cycles to a good approximation. At lower noise strengths, the theory becomes increasingly more accurate. The batch size required to obtain a pre-set degree of agreement between simulations theory will generally increase the closer a given choice of parameters is to the Hopf bifurcation, similar to what has been seen in \cite{Boland2008}. Although the mixed strategy profiles of individual players will evolve differently in any one run of the learning dynamics, there are statistically equivalent when an average over sufficiently many runs is taken. Power spectra such as those shown in Fig. \ref{fig:ps_ocomp} in particular are identical for different players, due to the symmetry of the well-mixed setup with respect to permutations of players. As we will see below, this is no longer the case in networked systems when different players have different connectivities.

\begin{figure}
\vspace{2em}
\centering
\includegraphics[scale=0.5]{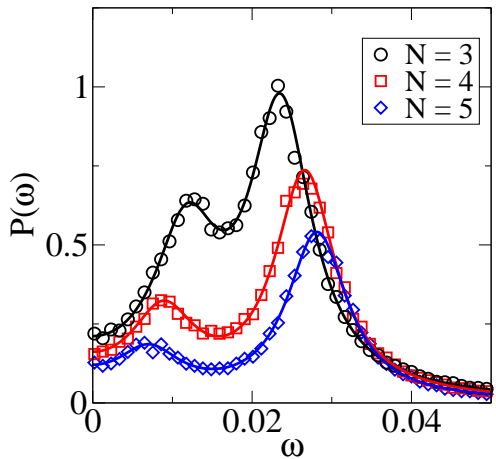}
\caption{\label{fig:ps_Ncomp} This figure shows the power spectrum for a player's contributory strategy at $\lambda = 0.005$, $\beta = 0.1$ for different numbers of players. The different colours indicate different numbers of players as described in the legend. Symbols are from simulations run at $\Omega = 100$, averaged over in excess of $100$ independent runs. Solid lines are from the theory.}
\end{figure}

Fig.~\ref{fig:ps_Ncomp} shows a comparison between the spectrum for a three, four and five player game. Higher numbers of players lead to smaller amplitudes of oscillations, greater distance between the two peaks and an increase in the ratio of the heights of the peaks. The difference in amplitudes of fluctuations may here be due to the fact that there are more possible payoffs in between the maximum and the minimum payoff as the number of players is increased. Smaller jumps between payoffs may therefore reduce the effects of stochasticity, keeping all other model parameters fixed.

\begin{figure}[t!!!]
\vspace{2.5em}
\centering
\includegraphics[scale=0.5]{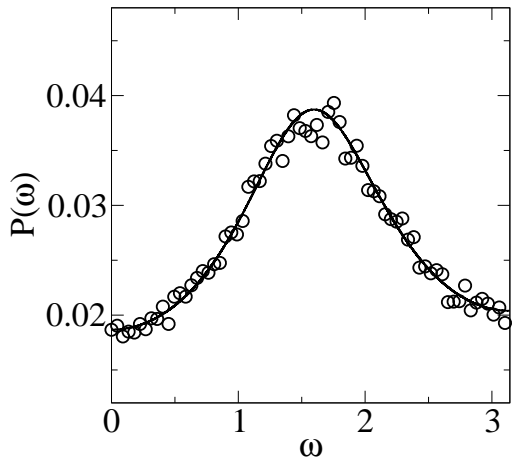}
\caption{\label{fig:expt} This figure shows a power spectrum for the contributing strategy which exhibits frequencies similar to those seen in experiments. The parameters for this five player game are $\Omega = 100$, $\beta = 2$, $\lambda = 1$.}
\end{figure}

Experimental situations often observe oscillations with much higher frequencies than the examples given here so far. An experiment by Milinski et al~\cite{Semmann2003}, although constructed differently to the situation this model mimics, shows oscillations of frequency $\omega \approx 2$. Fig.~\ref{fig:expt} shows an example with a high learning rate and rapid memory loss that reproduces such frequencies. We stress though that no claim is made that such memory-loss rates or intensity of choice are necessarily realistic. Experimental data for other games is available in \cite{Ho2007}. Instead the purpose of Fig. \ref{fig:expt} is mainly to show that the characteristic frequency of the observed quasi-cycles can vary over a wide range.

\section{Public good games on networks}
\label{sec:networks}

In real social systems populations are not well-mixed, instead people will only interact with certain others. It is no surprise that the study of games on networks, and the properties of networks in general, has become so popular, see e.g. \cite{Newmanbook, Barabasibook, Lieberman2005, Szabo2006}. Analytical progress in studying games on networks has been made \cite{Tarnita2009, Fu2009, Gross2005}, these methods typically assume meta-population models, with a large group of player placed at each node of the underlying network. We here take a different approach, and consider single individuals, connected via a static network. Within the game learning approach we can then apply the expansion outlined in Sec.~\ref{sec:stoc}, treating the strategies on each node as separate variables. The network structure is reflected in the calculation of the average payoffs and in the correlations of the noise. This approach produces a dynamical system whose dimensionality increases linearly with the number of nodes, making the study of large networks difficult although technically possible. However, up to recently, experimental situations are usually also restricted to small numbers of players \cite{Semmann2003, Camererbook, Ho2007}, with some notable exceptions, e.g. \cite{Grujic2010}.

\subsection{Batch-size expansion on a network}

In the networked model we assume that each player at each iteration of the game chooses a move ($C,D$ or $L$), according to his or her mixed strategy profile. Payoffs are then determined by Eq. (\ref{eq:payoff}), where the group of opponents faced by player $i$ is the set of his neighbors $\partial i$. The calculation then proceeds along similar steps to that of the well-mixed case. If there are $N$ nodes (players) in the network, and if each one of them chooses between $S$ pure strategies, then the resulting dynamics has $N(S-1)$ degrees of freedom, as before. In the limit of infinite batch size this leads to an $N(S-1)$ dimensional deterministic dynamical system in discrete time. As one crucial difference to the well mixed case the permutation symmetry between players is no longer present, even for homogeneous initial conditions. For general networks, different players will typically have different degrees, i.e., they face different numbers of opponents. Hence the mixed strategies played at deterministic fixed points will generally vary across the set of players (as before we here focus on the case of parameters in which deterministic learning converges). Carrying out the expansion in the inverse batch size one also finds that the network structure affects the noise correlators and the structure of the Jacobian of the deterministic dynamics. It is worth noting that we do not necessarily expect to see a breaking of the permutation symmetry for regular networks, i.e., when all players have the same degree $k$. For homogeneous initial conditions the mixed strategies of all players will then evolve identically under the deterministic dynamics. For heterogenous initial conditions and/or noisy dynamics this may not necessarily be the case though, as shown for the two-player case in \cite{Gomez2011}.

In the networked setting the expected payoff for player $i$ is given by
\be
\mu_{i,s} = \displaystyle \sum_{\bm{s}_{\partial_i}}  p_{i,\bm{s}_{\partial_i}}u(s,\bm{s}_{\partial_i})
\label{eqn:networkmu}
\ee
where $\bm{s}_{\partial_i}\in\{1,\dots,S\}^{|\partial i|}$ are the actions of those players connected to player $i$ (in the fully connected case this is simply the set of all other players), and where $p_{i,\bm{s}_{\partial_i}}$ is the probability for their joint action  $\bm{s}_{\partial_i}$ to occur. The analog of Eqs.~(\ref{eqn:samecorr}) and (\ref{eqn:diffcorr}) can be written as
\BE
&&\langle\widetilde{\xi}_{i,s}\widetilde{\xi}_{j,s^\prime}\rangle\nonumber \\ & =& \displaystyle\sum_{\bm{s}_{\partial_i\cup\partial_j}}  p_{\partial i\cup\partial j,\bm{s}_{\partial i \cup \partial j}}(u(s,\bm{s}_{\partial_i}) - \mu_{i,s})\\\nonumber&&~~~~~~\times(u(s^\prime,\bm{s}_{\partial_j}) - \mu_{j,s^\prime}), \label{eqn:networkcorr}
\EE
where $\partial i\cup \partial j$ is the union of the sets of neighbours of $i$ and of $j$.  The remaining steps are as in Sec.~\ref{sec:stoc}, with the appropriate modifications to take account of the changed payoff structure in the networked arrangement, Eq.~(\ref{eqn:networkmu}).
\subsection{Effect of degree}

\begin{figure}[t!!!]
\vspace{2.5em}
\centering
\includegraphics[scale=0.4]{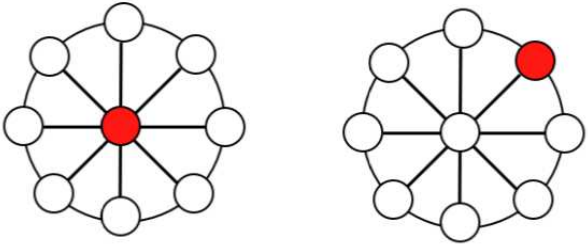}
\caption{\label{fig:networks}The network chosen for analysis. We will look at two players: the `hub' player (left) and the `ring' player (right).}
\end{figure}

We will restrict our discussion to the network shown in Fig.~\ref{fig:networks}, and we investigate the dynamics both for the central player, and for those on the outer spokes. The choice of this particular shape of network is to a certain extent arbitrary, it is important to keep in mind though that our theoretical approach applies to general graphs. The choice made here is hence mainly for purposes of illustration.

\subsubsection{Deterministic features}

\begin{figure}[t!!!]
\vspace{2.5em}
\centering
\includegraphics[scale=0.5]{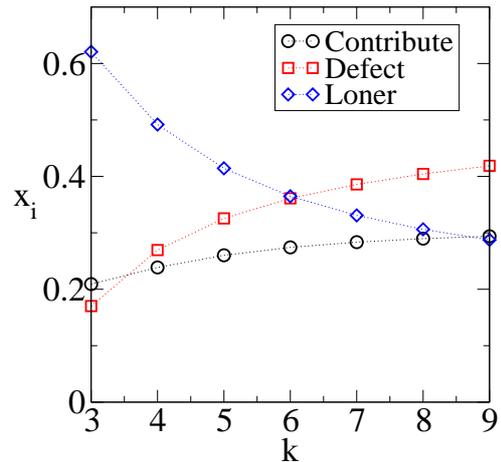}
\caption{\label{fig:detdeg} This figure shows the concentrations of the three strategies for the central player in networks of the type shown in Fig. \ref{fig:networks}, as a function of the player's degree $k$. The parameter values are $\beta = 0.1$, $\lambda = 0.005$. Results are from the deterministic map, lines connecting the symbols are guides to the eye.}
\end{figure}

\begin{figure}[t!!!]
\vspace{2.5em}
\centering
\includegraphics[scale=0.5]{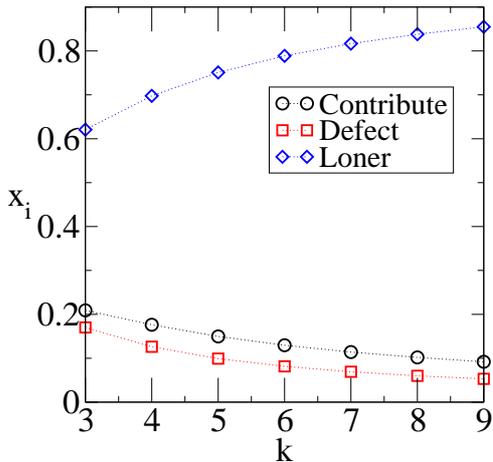}
\caption{\label{fig:detndeg} This figure illustrates how the mixed strategy profile of a player on the outer ring of networks of the type shown in Fig. \ref{fig:networks} changes, as the degree $k$ of the central player is varied. The parameter values are $\beta = 0.1$, $\lambda = 0.005$. Results are from the deterministic map, lines connecting the symbols are guides to the eye.}
\end{figure}

Let us first examining the effect of degree on the deterministic behaviour of the system, we focus on the fixed point regime, i.e., sufficiently quick memory loss. As discussed above the mixed strategies played by different players at deterministic fixed points will generally depend on their degree and the degree of their neighbours. To illustrate this we have considered networks of the type shown in Fig. \ref{fig:networks}, varying the degree $k$ of the central player. The resulting mixed strategy profile at convergence of the deterministic dynamics is shown in Fig.~\ref{fig:detdeg}  for the central player, and in Fig. \ref{fig:detndeg} for a player on the outer ring of the network. As the degree of the central players increases, so does the probability that he or she defects. For the outer player it is mostly the probability of abstaining (i.e., to play `loner') which increases, as the connectivity of the central player is increased. This is what one would expect given that their central neighbour is becoming increasingly likely to defect.

\subsubsection{Stochastic features}

\begin{figure}[t!!!]
\vspace{2.5em}
\centering
\includegraphics[scale=0.5]{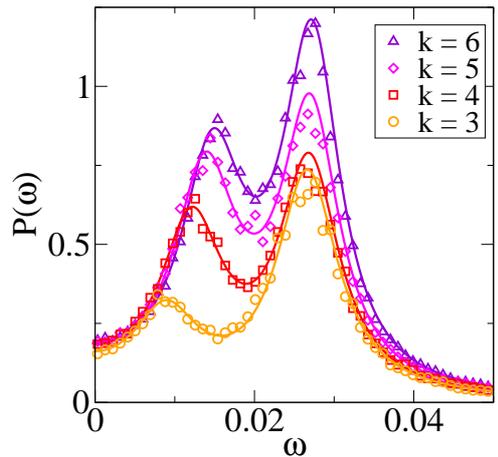}
\caption{\label{fig:stocdeg} Shown are the power spectra for the contributing strategy of a player in the centre of a ring-and-hub network. Different colours represent different degrees as shown in the legend. The parameter values are $\beta = 0.1$, $\lambda = 0.005$. Simulations are from $500$ runs at $\Omega = 10$.}
\end{figure}

\begin{figure}[t!!!]
\vspace{2.5em}
\centering
\includegraphics[scale=0.5]{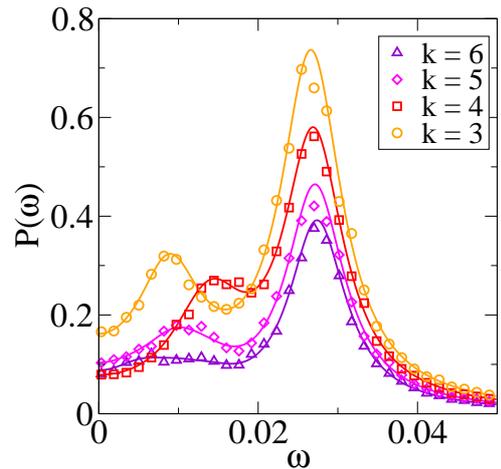}
\caption{\label{fig:stocndeg}The power spectra for the contributing strategy on the edge of a ring-and-hub network. Different colours represent different degrees of the `hub' neighbour. The parameter values are $\beta = 0.1$, $\lambda = 0.005$. Simulations are from $500$ runs at $\Omega = 10$.}
\end{figure}

As a final part of the analysis we study the effect of connectivity on oscillations induced by intrinsic noise in the learning process. In Fig. \ref{fig:stocdeg} we depict the power spectra of fluctuations of the probability with which the central player in our sample network plays strategy $C$. An analogous plot for a player on the outer ring of the network is shown in Fig. \ref{fig:stocndeg}. In both cases a good match between theory and simulations is found, even at relatively moderate batch sizes of $\Omega=10$. The theoretical approach based on a batch-size expansion is therefore successful in predicting the characteristic frequency of the observed quasi-cycles. Comparison of Figs. \ref{fig:stocdeg} and \ref{fig:stocndeg} with Fig. \ref{fig:ps_Ncomp} reveals an intriguing behaviour: (i) increasing the overall connectivity in the `hub-and-ring' network seems to reduce fluctuations of mixed strategies of players on the outer ring of the network. This is similar to the effect we have seen in the well-mixed case (see Fig. \ref{fig:ps_Ncomp}), where fluctuations are suppressed as the number of players in the group (or equivalently the total number of links in the graph) increases. In-line with this behaviour we find that the player on the outer ring of the network is more likely to abstain as the degree of the central node is increased, similar behaviour is found in the well-mixed case as the overall number of players is increased (not shown); (ii) The central player in the network however shows a different type of behaviour as their degree is increased. He or she is less likely to abstain (see Fig. \ref{fig:detdeg}), and fluctuations of the central player's mixed strategy increase with increasing degree (Fig. \ref{fig:stocdeg}). 

Care needs to be taken though in making direct comparisons between the network shown in Fig. \ref{fig:networks} and the well-mixed case. Increasing the degree $k$ of the central player in the network does not imply an increased connectivity of the players on the outer ring, and the situation is therefore different from that of a regular random network with uniform degree across all nodes. The juxtaposition between the behaviour of the central agent in the network and that of a player in the fully connected case indeed suggests that it is not only the degree of a player itself that determines his or her mixed strategy and the magnitude of fluctuations about it, but that on the contrary the degrees of the players he or she is connected to play an important role as well.

\section{Conclusion and outlook}
\label{sec:conclusion}

In summary we have investigated the deterministic and stochastic learning dynamics in multi-player games. Our approach builds on models from behavioural game theory \cite{Camerer2003,Ho2007}, and is similar to the Sato-Crutchfield formulation of game learning  \cite{Sato2002,Sato2002b, Sato2005}. The approach by Sato et al is however based on deterministic ordinary differential equations, and ignores effects of noise on the outcome of learning. Such stochastic effects have recently been studied for two-player games in \cite{Galla2009,Galla2011}, the main contribution of the current work is to extend these approaches to the case of multi-player games, and to games on networks. We have here successfully carried out expansions in the inverse noise strength. On the deterministic level Sato-Crutchfield learning, and its extensions to discrete time, exhibit features similar to those of the replicator, or replicator-mutator equations. On a stochastic level we confirm the presence of quasi-cycles in a wide range of model parameters, in particular stochastic learning can exhibit persistent oscillations in parameter regimes where deterministic learning converges. Based on the expansion in the inverse noise strength we are able to make analytical predictions on the existence or otherwise of quasi-cycles as a function of parameters of the learning dynamics and of the underlying game or network structure.

We believe that taking a modelling approach based on adaptation dynamics intrinsic to individual players, inspired by models of behavioural game theory, has a number of advantages over population-based approaches repying on birth-death processes. Learning models allow the incorporation of memory loss which has been shown experimentally to be an important factor. By contrast, since the replicator approach models the evolution of strategies rather than the behaviour of players it is often not straightforward to include such psychological or cognitive effects directly.  On the other hand, both approaches are not mutually exclusive, and there are indeed a number of similarities between them. Memory loss and mutation for example play similar roles in the respective modelling frameworks. Future research may therefore focus on drawing further analogies, and to use the existing knowledge on evolutionary models in finite populations to elucidate the mathematical structures and outcomes of stochastic learning models in further detail.

\begin{acknowledgments} 
AJB acknowledges an EPSRC studentship. TG is supported by an RCUK Fellowship (RCUK reference EP/E500048/1), and by the Engineering and Physical Sciences Research Council (EPSRC, references EP/I005765/1 and EP/I019200/1). The authors would like to thank Yuzuru Sato for useful discussions.
\end{acknowledgments}

\appendix
\section{Batch-size expansion for a well-mixed, three-player game}
As an example of the application of the equations derived in Sec.~\ref{sec:stoc}, we detail the steps involved in the calculation for the well-mixed case with $N=3$ and $S=3$. The average payoff for strategy $s$ used by player $i$ then becomes
\be
\mu_{i,s} = \sum_{s^\prime s^{\prime\prime}}a_{ss^\prime s^{\prime\prime}}x_{j,s^\prime}x_{k,s^{\prime\prime}},
\label{eqn:3mu}
\ee
where $j$ and $k$ indicate the other two players and $a_{ss^\prime s^{\prime\prime}}$ is an $3$-dimensional tensor containing the payoff for strategy $s$ when the other players use strategies $s^\prime$ and $s^{\prime\prime}$. Average payoffs for the other players are found through cyclic permutation of $i$, $j$ and $k$. Evaluating Eq.~(\ref{eqn:samecorr}) we find
\BE
&&\langle \widetilde{\xi}_{i,s}(\tau)\widetilde{\xi}_{i,s^\prime}(\tau')\rangle \nonumber \\
&=&\frac{\delta_{\tau\tau'}}{\Omega}\sum_{s'',s'''}x_{i,s''}x_{i,s'''}(a_{ss''s'''} - \mu_{i,s})\nonumber\\
&&~~~~~~~~~~~~\times (a_{s^\prime s''s'''} - \mu_{j,s^\prime}).\label{eq:s}
\EE

From Eq. (\ref{eqn:diffcorr}) we find  
\BE
\langle \widetilde{\xi}_{i,s}(\tau)\widetilde{\xi}_{j,s^\prime}(\tau')\rangle &=&\frac{\delta_{\tau\tau'}}{\Omega}\sum_{s'',s''',s''''}x_{i,s''}x_{j,s'''}x_{k,s''''} \nonumber\\
&\times& (a_{ss'''s''''} - \mu_{i,s})(a_{s^\prime s''s''''} - \mu_{j,s^\prime})\nonumber \\
\label{eq:d}
\EE
Eq.~(\ref{eqn:diffcorr}) differs from the two player-case studied in \cite{Galla2009,Galla2011} where the correlations of the noise associated with strategies belonging to different players is zero. In the three-player case, this noise is correlated through the actions of the third player.
If we introduce the ansatz $x_{i,s} = \bar{x}_{i,s} + \widetilde{x}_{i,s}$ as in Sec.~\ref{sec:stoc}, the average payoffs given by Eq.~(\ref{eqn:3mu}) can be separated into a term of order $\Omega^0$ and a term of order $\Omega^{-1/2}$

\be
\bar{\mu}_{i,s} = \sum_{s^\prime s^{\prime\prime}}a_{s s^\prime s^{\prime\prime}}\bar{x}_{j,s^\prime}\bar{x}_{k,s^{\prime\prime}}
\ee
\be
\widetilde{\mu}_{i,s} =  \sum_{s^\prime s^{\prime\prime}}a_{s s^\prime s^{\prime\prime}}(\bx_{j,s^\prime}\tx_{k,s^{\prime\prime}} + \tx_{j,s^\prime}\bx_{k,s^{\prime\prime}}).
\ee
The expansion of Eq.~(\ref{eqn:start}) for the three player case then gives the following for terms of order $\Omega^{-1/2}$.
\BE
&&\tx_{i,s}(\tau+1) - \tx_{i,s}(\tau) = \sum_{s^\prime}\left(\left.\pd{g_{i,s}}{\widetilde{x}_{i,s^\prime}}\right)\right|_{\widetilde {\bm x}=\widetilde{\bm \xi}=0}\widetilde{x}_{i,s^\prime} \\ \nonumber
&&+ \sum_{s^\prime}\left(\left.\pd{g_{i,s}}{\widetilde{x}_{j,s^\prime}}\right)\right|_{\widetilde {\bm x}=\widetilde{\bm \xi}=0}\widetilde{x}_{j,s^\prime} + \sum_{s^\prime}\left(\left.\pd{g_{i,s}}{\widetilde{x}_{k,s^\prime}}\right)\right|_{\widetilde {\bm x}=\widetilde{\bm \xi}=0}\widetilde{x}_{k,s^\prime} \\ \nonumber
&&+ \sum_{s^\prime}\left(\left.\pd{g_{i,s}}{\txi_{i,s^\prime}}\right)\right|_{\widetilde {\bm x}=\widetilde{\bm \xi}=0}\txi_{i,s^\prime}.
\EE
For player $i$, the coefficients multiplying $\txi_{j,s^\prime}$ and $\txi_{k,s^\prime}$ vanish. The remainder of the calculation then proceeds exactly as in Sec.~\ref{sec:stoc}.

\end{document}